\documentclass[12pt]{article}
\usepackage{amsfonts,bm,bbm}
\usepackage{mathrsfs}
\usepackage{amsmath}	
\usepackage{cite}
\usepackage{graphicx}
\usepackage{rotating}
\usepackage{hyperref}
\usepackage{verbatim}

\csname @addtoreset\endcsname{equation}{section}
\newcommand{\beq}{\begin{equation}}
\newcommand{\eeq}{\end{equation}}
\newcommand{\bea}{\begin{eqnarray}}
\newcommand{\eea}{\end{eqnarray}}
\newcommand{\ba}{\begin{array}}
\newcommand{\ea}{\end{array}}
\newcommand{\bit}{\begin{itemize}}
\newcommand{\eit}{\end{itemize}}
\newcommand{\nn}{\nonumber}

\newcommand{\mezzo}{\frac{1}{2}}
\newcommand{\complesso}{{\ \hbox{{\rm I}\kern-.6em\hbox{\bf C}}}}
\newcommand{\reale}{{\hbox{{\rm I}\kern-.2em\hbox{\rm R}}}}
\newcommand{\1}{ \,  \raisebox{+0.14em}{{\hbox{{\rm \scriptsize ]}} \raisebox{-0.2em}{\kern-.8em\hbox{1}}}} \, }  

\newcommand{\p}{\partial}

\renewcommand{\a}{\alpha}

\renewcommand{\d}{\delta}

\renewcommand{\l}{\lambda}
\renewcommand{\L}{\Lambda}

\newcommand{\m}{\mu}

\newcommand{\n}{\nu}
\renewcommand{\r}{\rho}
\newcommand{\s}{\sigma}

\renewcommand{\t}{\theta}

\begin{document}

\begin{titlepage}
\begin{flushright}
CECS-PHY-10/15
\end{flushright}
\vspace{1.5cm}
\begin{center}
\renewcommand{\thefootnote}{\fnsymbol{footnote}}
{\Large \bf Accelerating black hole in 2+1 dimensions}
\vskip 5mm
{\Large \bf and 3+1 black (st)ring}
\vskip 30mm
{\large {Marco Astorino\footnote{marco.astorino@gmail.com}}}\\
\renewcommand{\thefootnote}{\arabic{footnote}}
\setcounter{footnote}{0}
\vskip 10mm
{\small \textit{
Instituto de F\'{\i}sica,\\ 
Pontificia Universidad Cat\'olica de Valpara\'{\i}so \\
and\\
Centro de Estudios Cient\'{\i}ficos (CECS), Valdivia,\\ 
Chile\\}
}
\end{center}
\vspace{3.8 cm}
\begin{center}
{\bf Abstract}
\end{center}
{A C-metric type solution for general relativity with cosmological constant is presented in 2+1 dimensions. It is interpreted as a three-dimensional black hole accelerated by a strut. Positive values of the cosmological constant are admissible too.\\ Some embeddings of this metric in the 3+1 space-time are considered: accelerating BTZ black string and a black ring where the gravitational force is sustained by the acceleration.}
\end{titlepage}

\section{Introduction}

Beside the fact that general relativity in three space-time dimensions is trivial, because the lack of dynamical degrees of freedom, still it admits black holes solutions. In fact the static and rotating black hole in 2+1 dimension is very well known as remarkably discovered by Ba\~nados, Teitelboim and Zanelli (BTZ) in \cite{bhtz}. Also well known is the charged and the electro-rotating flavours \cite{Qbtz}. Taking inspiration from the four dimensional C-metric (for references see \cite{podo}, \cite{lemos} and \cite{int-c}), we are here interested to apply acceleration to the three dimensional black hole (section \ref{accBTZ}). This is done with the motivation that such acceleration, and the physical object that produce it, could play an important role in 3+1 general relativity solutions with non-trivial topology, as shown in section \ref{3+1}. Our supposition is based on analogy with the five dimensional case, as explained in \cite{empa08}: to forge a black ring one has to bend a string built with the cross product of a line and a Schwarzschild black hole, and then balance the gravitational self attraction which makes the string self collide. If one would do so in four dimensions, the natural candidate to start with is the cross product of a line and the three dimensional black hole, thus the cosmological constant is necessary. Cosmological constant is also prominent in introducing a length scale to obtain different scales for the radius and the thickness of the ring. In \cite{empa08} the absence of black (st)rings in vacuum 4D general relativity is attributed to the lack of such a scale.

\section{Accelerating black hole in 2+1 dimensions}
\label{accBTZ}

We begin considering the Einstein-Hilbert action for standard general relativity, with generic cosmological constant $\L$, in three dimensions\footnote{We have set, for convenience, the gravitational constant $G=1/8$}:
\beq
         I[g_{\m\n}] = \frac{1}{2 \pi } \int d^3 x \sqrt{-g} (R - 2\L)
\eeq
Extremization of the action with respect the metric $g_{\m\n}$ yields the Einstein field equations:
\beq \label{einstein}
                       R_{\m\n}-\mezzo R \ g_{\m \n} + \L g_{\m\n} = 0
\eeq
Being inspired by the four dimensional C-metric \cite{int-c}, we start proposing a similar ansatz but in $2+1$ dimensions:
\beq
       ds^2 = \frac{1}{(1+\a r \cos \t)^2} \left[ -f(r) dt^2 + \frac{dr^2}{f(r)} + \frac{r^2 d\t^2}{g(\t)} \right]
\eeq
a general solution for (\ref{einstein}) in terms of the unknown functions $f(r)$ and $g(\t)$ can be find:
\beq \label{fg}
       f(r)= c_0 + c_1 \ r + c_2 \ r^2  \ \ ,  \qquad  g(\t)= \frac{c_0 \a^2 \cos^2(\t) - c_1 \a \cos(\t) + c_2 + \Lambda}{\a^2 (\cos^2(\t)-1)} 
\eeq
Where $c_0 , c_1 , c_2 $ are arbitrary integration constant. This metric describes locally a constant curvature space-time: $R^{\m\n}_{\ \ \r\s}=\L \ ( \d^\m_{\ \r}\d^\n_{\ \s}-\d^\m_{\ \s}\d^\n_{\ \r} )$.
In order to have a significant $\a \rightarrow 0$ limit we select a particular metric choosing, from (\ref{fg}), the following integration constants:
$$ c_0=1-m \quad , \qquad c_1 = 0 \quad , \qquad c_2 =  \a^2 (m-1) -\Lambda   $$
After rescaling the $\theta$ coordinate one obtains the solution:
\begin{align} \label{abtz}
\hspace{-1.0 cm}
     ds^2=\frac{1}{\left[1+\a r \cos \left(\theta \sqrt{1-m}\right)\right]^2} \bigg\{ &- \Big[ 1 - m + r^2 \big[ \a^2 (m-1) - \L \big] \Big] dt^2 \\
                            &+  \frac{dr^2}{1-m+r^2 \big[ \a^2 (m-1) - \L \big]} + r^2 d \t^2 \bigg\} \nonumber 
\end{align}
The coordinates $(t,r,\t)$ are chosen to be polar, so their range is $-\infty<t<\infty$, $r\geq 0$, $-\pi \leq \t \leq \pi$; since $\theta$ is an angular coordinate, the points $\t=-\pi$ and $\t=\pi$ can be considered identified. Other choises for the coordinate ranges and identification are possible and give rise to different spacetime geometries
; our choice is motivated by the will to model an accelerating 2+1 black hole. 
In fact when the parameter $\a$ is imposed to be null one has:
\beq \label{btz}
     ds^2 =  - \big(  1 - m - \L \ r^2  \big) dt^2 + \frac{dr^2}{ 1 - m - \L \ r^2 } + r^2 d\t^2 
\eeq
which is precisely the static BTZ black hole metric in the gauge where the ground state is the (A)dS space-time. We prefer this form of the metric respect to the one of \cite{bhtz} because when the mass\footnote{Computed respect to the (A)dS background, as done for instance in \cite{dete}} $m\rightarrow0$ one does recover from (\ref{btz}), not just an asymptotically (A)dS space-time as in \cite{bhtz}, but exactly the standard vacuum (A)dS space-time. Similarly this is what happens in 3+1 (or higher) dimension, for instance to the Schwarzschild-(A)dS black hole. 
The fact that in this gauge when $m \in [0,1)$ naked singularity occurs (while black holes for $m \in [1,\infty)$) it's a typical feature of Chern-Simons gravity or odd-dimensional Lovelock theories, such is three-dimensional general relativity.\footnote{Recently $m \in [0,1)$ states of the black hole spectrum were physically dignified as topological defects in \cite{zan-oli}, representing particles in $AdS_3$ (see also \cite{steif}) .}

The metric (\ref{abtz}) describes, for $m \ge 1$ an accelerating black hole whose event horizon is located at:
$$ r_h = \sqrt \frac{m-1}{\a^2 (m-1)-\L} $$

A physical interpretation to the $\a$ parameter can be given, following the argument of \cite{podo} and \cite{lemos}. Usually, as for the 3+1 C-metric, one considers the weak field limit of the solution (\ref{abtz}), that is $m\rightarrow0$:
\beq \label{weak}
       ds^2=\frac{1}{\big[1+\a r \cos (\theta )\big]^2} \left\{ - \big[ 1 - r^2 ( \a^2 + \L ) \big] dt^2 + \frac{dr^2}{1-r^2 ( \a^2 + \L )} + r^2 d \t^2 \right\}  
\eeq
The weak field limit is usefull because, in this case, black holes can be considered as test particles and cease to deform the spacetime and inertial frames around them.\footnote{A metric similar to (\ref{weak}) is studied by \cite{anber} in the case of null cosmological constant.}
The 3D timelike worldlines $x^\m(\l)$ of an observer with $r=\bar{r}=constant$ and $\theta=0$ can be obtained by the property $u_\m u^\m=-1$ of the 3-velocity defined by $u^\m=dx^\m/d\l$:
$$   x^\m(\l)=\left[\frac{1+\a \bar{r} \cos(\theta)}{\sqrt{1-\bar{r}^2(\a^2 + \L)}}\l,\bar{r},0 \right] \ \ \ ; $$
where $\l$ is the proper time of the observer. Then the magnitude $a$ of the 3-acceleration, $a^\m=(\nabla_\n u^\m) u^\n$, for this kind of observer results
\beq   \label{acc}
                    |a| = \sqrt{a_\m a^\m} \ \Big|_{\bar{r}=0} = \a
\eeq
Since $a_\m u^\m=0$, the value $|a|$ is also the magnitude of the 2-acceleration in the rest frame of the observer. From eq. (\ref{acc}) we achieve the conclusion that the origin of the metric (\ref{weak}), $\bar{r}=0$ is being accelerated with an uniform acceleration whose value is precisely given by the constant $\a$, so (\ref{weak}) is nothing but the accelerating (A)dS space-time. Outside of the weak field limit similar results can also be obtained: $a_\m a^\m \ |_{\bar{r}=0} = \a^2 \ (1-m)$ for $m\neq1$, while $a_\m a^\m \ |_{\bar{r}=0} = -\L $ for $m=1$. \\
Thus now we can interpret the BTZ metric (\ref{btz}) as the non-accelerating limit of the (\ref{abtz}). Thanks to this limit the parameter $m$ which appears in (\ref{abtz}) can be interpreted as the mass parameter of our solution. Of course in case of non null acceleration this does not coincide with the mass of the metric (\ref{abtz}), but is somehow related to the mass. Actually is not clear how to compute energy in this class of accelerating space-time, neither in standard 3+1 dimensions, because of the non trivial asymptotic. So usually in the literature (see \cite{int-c}) the mass value is estimated by thermodynamical argument as follows: The first law of black hole thermodynamics states that $ dM=T dS = \frac{k}{8\pi G} dA $
; while the surface gravity $k$ is defined by $ k^2=-\mezzo \nabla^\m \chi^\n \nabla_\m \chi_\n $ where $\chi^\m$ is the killing vector $\p_t$:
$$  k=r_h \left[ (m-1) \a^2 -\L  \right] \bigg|_{m=1+\frac{\L r_h^2}{r_h^2 \a^2-1}} = \frac{\L r_h}{r_h^2 \a^2-1}  $$
The area of the accelerating black hole horizon $A$ is given by:
\beq \label{area}
      A = \int_{-\pi}^\pi \sqrt{g_{\t\t}} \> \bigg|_{\!\!\begin{array}{l}\scriptscriptstyle r=r_h\\[-1.2ex]\scriptscriptstyle  t=\hbox{\tiny const.}\end{array}}\!\!\!\! \ d\t \ = \ \frac{4}{\sqrt{\L}} \arctan \left[ \sqrt{\frac{\a r_h -1}{\a r_h + 1}} \tanh \left( \frac{\pi}{2} \sqrt{\frac{\L r_h^2}{r_h^2 \a^2 -1}} \right) \right]
\eeq
So the mass estimation is, up to a integration constant:  
\beq
 M = \int dM = \int \frac{2 \sqrt{\L} }{\pi (\a^2r_h^2-1)^{3/2}}  \frac{ \a r_h \sinh \left(\pi\sqrt{\frac{\L r_h^2}{\a^2 r_h^2-1}}\right) - \pi \sqrt{\frac{\L r_h^2}{\a^2 r_h^2-1}} }{1+\a r_h\cosh \left( \pi\sqrt{\frac{\L r_h^2}{\a^2 r_h^2-1}}\right)} dr_h \nn
\eeq
For small values of the acceleration parameter $\a << 1$, thus, in practice we restrict only to the $\L<0$ sector, this integral can be evaluated:
\beq
     M = -\L r_h^2 + 2 \a \left[ \frac{r_h}{\pi^2} \cos \left( \pi r_h \sqrt{\L} \right) + \frac{\L\pi^2r_h^2-1}{\sqrt{\L}\pi^3} \sin \left( \pi r_h \sqrt{\L} \right) \right] + O(\a^2) \nn
\eeq
Evidently in case of null acceleration this results is coherent with the BTZ one: $M[r_h]=m[r_h]+k_0$. Note that the integration constant $k_0$ is background dependent, $k_0=0$ for the AdS background. \\
The main difference of this three dimensional accelerating black hole, compared with the four dimensional C-metric, is the absence of a pure acceleration horizon. In fact the effect of the acceleration merges with the cosmological constant pressure to give an unique event horizon. This happens because also the standard three dimensional event horizon is just that of an accelerated observer, until one identifies $\theta$ as an angular coordinate \cite{bhtz}. The peculiar combination between $\L$ and $\a$ has the remarkable consequence that also positive values for the cosmological constant are admissible in order to get a black hole configuration. This is something unexpected since even a no-go theorem (see \cite{ida}) is found to justify the lack of a 2+1 black hole, in standard general relativity, for positive cosmological constant. In our case the no-go theorem is circumvented in one of the hypothesis of regularity of the horizon: the horizon of the accelerating solution (\ref{abtz}) is continuous but not smooth as required in \cite{ida}. It is worth to note that the Riemannian curvature tensor remains locally constant everywhere, but in fact in $\theta=\pm \pi$ the metric is not differentiable because of an angular singularity, a common feature which characterises this kind of accelerating black holes. Here the singularity is not a conical one due to a standard deficit angle as the four dimensional case, this may appear as another difference. But it is just because otherwise in one dimension less there would be no room for a strut or a string.
\begin{figure}[h]
\begin{center}
\includegraphics[angle=0, scale=0.45] {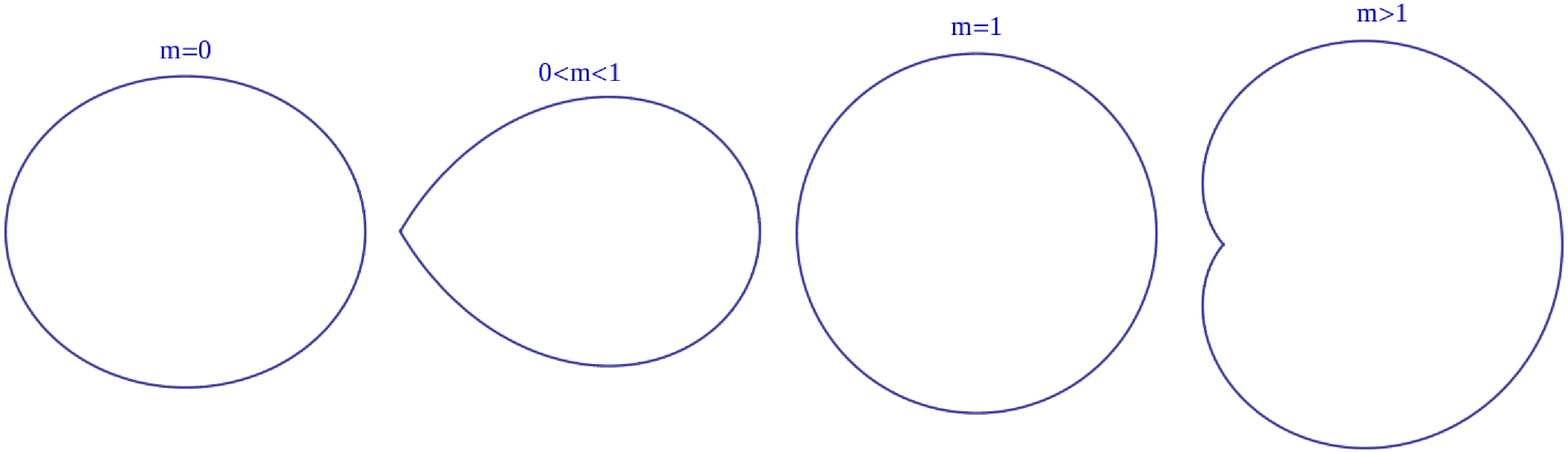}
\end{center}
\caption{\small{Polar plot ($\bar{r},\t$) of $\bar{r}=$constant radial curves embedded in the plane $\mathbb{E}^2$. When $\bar{r}=r_h$ they illustrate horizon deformations for various values of the mass: $m=0$, $0<m<1$, $m=1$ and $m>1$.}} 
\label{horizont}
\end{figure}
In figure \ref{horizont} are polar plotted the shapes of the horizon, for various values of the mass parameter $m$, as embedded in the two dimensional euclidean plane $\mathbb{E}^2$. Note that the sharp vertex occurring for $0<m<1$ and $m>1$ is an artifact of the embedding, in the real curved space-time there are no sharp vertexes. In fact the two vectors normal to the horizon in $\theta=\pi	$ and $\theta=-\pi$, denoted by $n^\m_{(\pm)}$, are parallels: $  n^\m_{(+)} \ n_{(-)\m} = 1  $.
Nevertheless the angular singularity in the conformal factor persist as pointed out by the undifferenciability of the metric in $\t=\pm \pi$ (for $m\neq 0,1$). It is usually associated to the presence of a semi-infinite cosmic string, or a strut, which is pulling the black hole along the $\t=\pi$ axis. In this picture the tension of the string $\tau$ is responsible for the acceleration of the 2+1 dimensional black hole. Thanks to the Israel junction condition it is possible to compute the strength of the strut's force, geometrically due by a jump in the extrinsic curvature $\mathcal{K}_{\m\n}$:
\begin{equation} \label{Tmn}
        \big[\mathcal{K}_{\m\n} \big]^{\t=\pi}_{\t=-\pi} - h_{\m\n} \big[ \mathcal{K} \big]^{\t=\pi}_{\t=-\pi}   = - \pi h_{\m\n} \tau   
\end{equation}
where $h_{\m\n}$ is the induced metric on the $\theta=\pm \pi$ surface, $\mathcal{K}=\mathcal{K}_{\m\n}h^{\m\n}$ and the tension is	
$$\tau= - 2 \a \  \sqrt{m-1} \   \sinh(\pi \sqrt{m-1}).$$
The negativity of $\tau$ indicate that we are dealing with a strut that is pushing the black hole, rather than a pulling string. As can be read from the metric (\ref{abtz}) or perceived from figure \ref{horizont}, the strut tension $\tau$ vanishes for $m=0,1$, where the metric acquire regularity.
The metric (\ref{abtz}) is supported by the surface stress-energy tensor that can be extract from (\ref{Tmn}) and, for sake of precision, it should be better added in the equation of motion (\ref{einstein}).\\
The casual structure and the Carter-Penrose diagram is similar to the BTZ one, although now the casual singularity in $r=0$ has an acceleration $\a$. 
\\
A rotating and accelerating metric can be obtained by an improper boost in the ($t,\theta$) plane of (\ref{abtz}) as explained in \cite{Qbtz}. Since in three dimensions there is no room for rotating around the strut axis as in four dimensions, the only way to pursue rotation is have the strut rotate with the black hole. As expected that metric reduces to the (\ref{abtz}) when the rotation is null, while it reduces to the rotating BTZ metric when the acceleration parameter vanish. \\
Since the metric (\ref{abtz}) remains finite for $r \rightarrow \infty$ one may also think to extend the spacetime also in the negative $r$ sector in order to reach the conformal infinity for $r=-1/(\a \cosh(\t \sqrt{m-1}))$. This can be done in general for $\L>0$. While for $\L<0$, depending also on the reciprocal values of the parameters $\L,m$ and the angular direction $\t$, one may encounter another killing horizon corresponding to the negative root of the $f(r)$ function before the conformal infinity. But here we are not interested in that. Instead we are more interested on the possible embeddings of this metric in the $3+1$ dimensional space time, as presented in the next section.\\



\section{Embeddings in 3+1}
\label{3+1}

\subsection{Accelerating BTZ black string}

The most direct embedding in 3+1 dimensions of the 2+1 solution (\ref{abtz}) is obtained in the spirit of \cite{empaII}. A string like object can be written by a warped product of the three dimensional metric and a line element $dz$:
\begin{align} \label{abtz-string}
\hspace{-0.5 cm}
    ds^2=\frac{\cos^2(z)}{\left[1+\a r \cos \left(\theta \sqrt{1-m}\right)\right]^2} \bigg\{ &- \left[ 1-m + r^2 \left( \a^2 (m-1) - \frac{\L}{3} \right) \right] dt^2  +  \\ 
   &   +  \frac{dr^2}{1-m+r^2 \left( \a^2 (m-1) - \frac{\L}{3} \right)} + r^2 d \t^2 \bigg\} \ + \ \frac{3}{\L} \  dz^2  \nn
\end{align}
To preserve the correct metric signature $(-,+,+,+)$ the z coordinate have to be rotated to the imaginary plane when dealing with negative cosmological constant: $z \rightsquigarrow iz$.
The standard BTZ black string of \cite{empaII} is precisely recovered from (\ref{abtz-string}) in the limit of $\a=0$ (of course in that case just negative cosmological constant can be considered).
But the periodic dependence on the $z$ coordinate, for positive $\L$ (and $\a \neq 0$), suggests that in this case the string singularity can be though of closed type. Thus the horizon's topology has a toroidal geometric structure, but with a couple of points in $z=\pm \pi/2$  where the horizon throat shrinks to zero\footnote{Alternatively is possible to achieve the spherical topology, for $\Lambda>0$, considering $-\pi/2<z<\pi$; the casual singularity would be a $\pi$-length segment along the $z$-axes.}.
Anyway many other features are shared between the static and accelerating black string. For instance the space-time curvature still remains trivially constant as the 2+1-dimensional case: $R^{\m\n}_{\ \ \r\s}=\L/3 \ ( \d^\m_{\ \r}\d^\n_{\ \s}-\d^\m_{\ \s}\d^\n_{\ \r} )$.\\
In this case the acceleration is provided by a two space dimensions membrane, which results a semi-infinite plane, at least for negative $\Lambda$.

\subsection{Black Ring}

Our main interest is in topological non-trivial solution, so we consider an ansatz with toroidal base manifold, finally. Not the one of constant curvature obtained by identification of a flat rectangle edges, but rather a doughnut embedded in the full 3+1 space-time, whose metric and curvature is giving, thinking $r$ and $t$ constant, as follows:
\beq 
    ds^2 = r^2 d\t^2 + \left[R_0+r\cos\left(\t \sqrt{1-m}\right)\right]^2 d\phi^2   \ , \qquad   R^{\t \phi}_{\ \ \t \phi}  = \frac{(1-m) \cos \left( \t \sqrt{1-m} \right)}{r\left[R_0 + r \cos \left( \t \sqrt{1-m} \right) \right]} \nn
\eeq
When $m\neq0\neq1$, but still $\t=\pm\pi$ identified, the circular section of the torus is deformed, respect to the smooth $m=0,m=1$ cases, in a drop or a cardioid shaped section (see figure \ref{horizont}). Using this base manifold with generic $m$ and the acceleration conformal factor of the previous section \ref{accBTZ} one can obtain a ring-like solution of Einstein equations (\ref{einstein}):
\bea \label{BRing}
     \hspace{-1.0 cm} 
     ds^2&=&\frac{1}{\left[1+\a r \cos \left(\theta \sqrt{1-m}\right)\right]^2} \bigg\{ - \left[ 1 - m + r^2 \left[ \a^2 (m-1) - \frac{\L}{3} \right] \right] dt^2 + \\
                            &+&  \frac{dr^2}{1-m+r^2 \left[ \a^2 (m-1) - \frac{\L}{3} \right]} + r^2 d \t^2 + \big[R_0 + r \cos(\t\sqrt{1-m}) \big]^2 d\phi^2 \bigg\} \nn
 \eea
where the bigger radius of the torus $R_0$ is related to the acceleration and mass parameter in this way: 
$$   R_0=\frac{\a(m-1)}{\a^2(m-1)-\L/3} = \a r_0^2  $$
It is evident that the acceleration plays a fundamental role in the topological structure of the solution: when the acceleration parameter decreases also the radius of the torus $R_0$ shrinks until it vanishes for $\alpha=0$; in the latter case local spherical symmetry in the base manifold is achieved (while globally one has a portion of $(A)dS_4$). So, from a physical point of view, the acceleration sustains and balances the gravitational attraction of the ring singularity, providing an equilibrium configuration (whose stability is unclear).
\begin{figure}[h]
\begin{center}
\includegraphics[angle=0, scale=0.8] {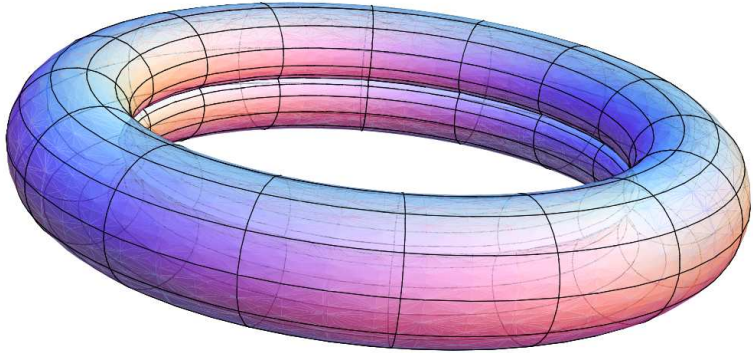}
\end{center}
\caption{\small{``Accelerating'' black ring horizon embedded in $E^3$} }
\label{toro}
\end{figure}\\
Considering a $\phi$-constant slice of (\ref{BRing}) one has exactly the accelerating black hole in 2+1 of section \ref{accBTZ}, modulo the dimensional rescaling of the cosmological constant.\\
The topology of the horizon, located at $r_0$, can be confirmed with the help of the Gauss-Bonnet theorem. It is  simple to compute the Euler characteristic when the mass parameter gives smooth 2-surface for constant time and radius, that is $m=0$ and $m=1$. For different values of $m$ one has to take into account the correction to the Gauss-Bonnet theorem due to the sharp edge, which is given by the Gibbons-Hawking term. This involves the jump on the trace of the extrinsic curvature $\mathcal{\bar{K}}$ and the induced metric $\bar{h}$ on the circular sharp edge. Consider the surface $\mathcal{S}$ described by the two dimensional metric obtained by (\ref{BRing}) fixing $r=\bar{r}=$const and $t=$const, whose embedding in the three dimensional euclidean space $E^3$ is portrayed in figure \ref{toro}:
$$ d\bar{s}^2 = \frac{1}{\left[1+\a \bar{r} \cos \left(\theta \sqrt{1-m} \right)\right]^2} \left\{ \bar{r}^2 d \t^2 + \big[R_0 + \bar{r} \cos(\t \sqrt{1-m}) \big]^2 d\phi^2 \right\} $$
Its Euler characteristic is null: 
\bea
      \chi(\mathcal{S})&=&\frac{1}{4\pi} \left( \int_{\mathcal{S}} \sqrt{\bar{g}} \ \bar{R} \ d\t \ d\phi \ + 2 \int_{-\pi}^\pi \left[\sqrt{\bar{h}} \  \bar{\mathcal{K}} \right]^{\t=\pi}_{\t=-\pi}  \ d\phi \right) = \nn \\ 
        &=& \frac{\a R_0-1}{4 \pi} \sqrt{1-m} \int_{-\pi}^\pi d\phi \int_{-\pi}^{\pi} \frac{\cos(\t \sqrt{1-m}) + \bar{r} \a}{[1+\a \bar{r} \cos(\t \sqrt{1-m})]^2} d\t  \nn \\ &-& \frac{\a R_0-1}{2 \pi} \int_{-\pi}^{\pi} \frac{2 \sin(\pi \sqrt{1-m})}{1+\a\bar{r}\cos(\pi \sqrt{1-m})} \ d\phi \ \  =  \ 0 \nn
\eea

so, since $\chi(\mathcal{S}) = 2 - 2g$, the genus of $\mathcal{S}$ is 1: toroidal topology $S^1 \times S^1$. Observe that the irregularity on the horizon can be cast also in the external part of the torus, as happens in the 5D static ring \cite{empa_real0}. \\
The causal singularity is located along $r=0$,  forming a circle in the $\phi$ direction. It's a naked singularity with a cosmological horizon for $m\in(0,1)$ while a black hole type for $m>1$. Coordinate $r$ is not a standard polar radius but rather the distance from the ring singularity.  Of course in order to have a proper black ring torus horizon the parameters of the metric are somewhat constrained to assure that the ring radius is larger than the ring thickness, $R_0 \geq r_0$: $0\leq \L \leq 3 (m-1) \a^2$. Note that in the null cosmological constant case we can have just a plump  horn ring with $R_0=r_0$, as expected in \cite{empa08}, because a lack of the lengh scale furnished by $\L$. It's not anymore possible tune the acceleration parameter $\a$ to make one radius arbitrary larger than the other one. While in the other limiting case, that is $\L=3(m-1)\a^2$, $R_0$ grows to infinity. To have a well behaved coordinate's set and an Hausdorff manifold, the range of $r$ have to be restricted when $\vert \t \vert>\pi/2$ to $r\ge-R_0/\cos(\t \sqrt{1-m})$. That fact is maybe clearer in toroidal coordinates (\ref{torcor}) but, on the other hand, this coordinates patch makes the physical interpretation of the metric (\ref{BRing}) more opaque. Anyway to get the the metric (\ref{BRing}) in the usual ring coordinate just rename $y=-1/\a r$, $x=\cos(\t\sqrt{1-m})$ and rescale time:
\bea  \label{torcor}
      ds^2=\frac{1}{\a^2 (y-x)^2} \bigg\{&-&\left[(y^2 -1) (1-m) - \frac{\L}{3 \a^2}\right] dt^2 + \frac{dy^2}{(y^2 -1) (1-m) - \frac{\L}{3 \a^2}}     + \nn \\                           &+& \frac{dx^2}{(1-x^2)(1-m)} + \left( \a R_0 \ y -x  \right)^2 d\phi^2  \bigg\}
\eea
This form of the solution may be of some utility for those people interested in finding a (A)dS ring five or in higher dimensions.

Overall this metric (\ref{BRing}) or (\ref{torcor}) is again locally (A)dS again despite the fact that the base manifold is not of constant curvature. However note that this behaviour is conceptually different from the topology of the accelerating black string (\ref{abtz-string}), where the toroidal topology is archived by means of the identification on the fourth coordinate, when $\L>0$. That identification forced a topology change in the whole universe, which is not here the case.

\section{Comments and Conclusions}
In this paper a C-metric type solution for 2+1 general relativity with cosmological constant is presented. It is analysed following the standard four-dimensional techniques: in the weak field approximation the extra parameter $\a$ is found to be the acceleration of an observer in the origin of coordinates. When $\alpha$ parameter vanishes the usual static BTZ black hole is recovered. Thus the metric (\ref{abtz}) is interpreted as an accelerating three-dimensional black hole, for certain range of the mass parameters: $m>1$. The acceleration is provided by a one-space-dimensional semi-infinite strut whose tension is proportional to the jump in the extrinsic curvature. For $0\le m<1$ the same solution represents an accelerating naked singularity with a cosmological horizon pulled by a string. A remarkable fact is that black hole configurations are admissible even for positive cosmological constant (whenever is smaller than a certain amount of acceleration: $\Lambda < \alpha^2 (m-1)$). This because the pushing effect of the strut arithmetically adds to the acceleration provided by the cosmological constant.\\
Moreover a couple of embeddings of this metric in the 3+1 dimension are considered. First an accelerating (by a two space dimension membrane) black string, whose horizon topology depends on the sign of the cosmological constant: cylindrical or toroidal for negative or positive $\Lambda$ respectively. Again the standard BTZ black string is retrieved when $\alpha=0$. Lastly it is proposed a ring singularity covered by a toroidal horizon where the gravitational force is balanced by the acceleration supplied by a disk (or puncured plane in case of external irregularity) that, in fact, sustains the ring. Each $\phi$-constant slice represents an accelerating black hole of one dimension less. Up to the author knowledge this is, though not regular, the first analytical black ring solution in four dimensions and also the first not asymptotically flat in any dimensions. \\
For future perspective would be very interesting counterweight the gravitational attraction, instead of the acceleration only, by a smoother centrifugal effect due to the angular momentum. This may provide a regular horizon such as been done in five dimension by Emparan and Reall passing from an irregular static ring in \cite{empa_real0} to a regular rotating one in \cite{emp-real}. The case of null acceleration implies only negative cosmological constant, so the Hawking topology censorship theorem can be avoided. Nevertheless the presence of the cosmological constant preclude the use of standard generating solutions techniques.

\section*{Acknowledgements}
\small
I would like to thank Eloy Ayon Beato, Fabrizio Canfora, Fiorenza de Micheli, Dietmar Klemm, Julio Oliva, Souya Ray, David Tempo, Steve Willison and Jorge Zanelli for fruitful discussions. I'm deeply indebted to Hideki Maeda for his continuous encouragement, suggestions and comments, without his help this work just wouldn't have risen.\\
\small This work has been partially funded by the Fondecyt grant 1100755 and by the Conicyt grant ``Southern Theoretical Physics Laboratory'' ACT-91. The Centro de Estudios Cient\'{\i}ficos (CECS) is funded by the Chilean Government through the Centers of Excellence Base Financing Program of Conicyt. CECS is also supported by a group of private companies which at present includes Antofagasta Minerals, Arauco, Empresas CMPC, Indura, Naviera Ultragas and Telef\'{o}nica del Sur.

\end{document}